\documentclass[default, notitlepage]{ml4h-jnl}% Default
\jyear{2024}%

\theoremstyle{thmstyleone}%
\theoremstyle{thmstyletwo}%

\theoremstyle{thmstylethree}%

\usepackage{hyperref}
\usepackage{subfigure}

\begin{document}

\title{Pre-trained protein language model for codon optimization}

%% GivenName	-> \fnm{Kaushik.}
%% Particle	-> \spfx{Manjunatha} -> surname prefix
%% FamilyName	-> \sur{Upadhya}
%% \author*[1,2]{\pfx{Dr} \fnm{Kaushik} \spfx{Manjunatha} \sur{Upadhya} \sfx{IV} \tanm{Poet Laureate} 

\author*[1]{\fnm{Shashank} \sur{Pathak}}\email{spathak2@ualberta.ca}

\author[*1]{\fnm{Guohui} \sur{Lin}}\email{guohui@ualberta.ca}


\affil*[1]{\orgdiv{Department of Computing Science}, \orgname{University of Alberta}, \city{Edmonton}, \state{Alberta}, \country{Canada}}

\affil*[2]{\orgdiv{Department of Computing Science}, \orgname{University of Alberta}, \city{Edmonton}, \state{Alberta}, \country{Canada}}

% \affil[3]{\orgdiv{Department}, \orgname{Organization}, \orgaddress{\street{Street}, \city{City}, \postcode{610101}, \state{State}, \country{Country}}}


% \abstract{The abstract serves both as a general introduction to the topic and as a brief, non-technical summary of the main results and their implications. Authors are advised to check the author instructions for the journal they are submitting to for word limits and if structural elements like subheadings, citations, or equations are permitted.}

%%================================%%
%% Sample for structured abstract %%
%%================================%%

\abstract{
\paragraph*{\textbf{Motivation}} Codon optimization of Open Reading Frame (ORF) sequences is essential for enhancing mRNA stability and expression in applications like mRNA vaccines, where codon choice can significantly impact protein yield which directly impacts immune strength. 
In this work, we investigate the use of a pre-trained protein language model (PPLM) for getting a rich representation of amino acids which could be utilized for codon optimization.
This leaves us with a simpler fine-tuning task over PPLM in optimizing ORF sequences.
% Our model, trained on hg19 human genes filtered by sequence length and minimum free energy (MFE). 
\paragraph*{\textbf{Results}}
The ORFs generated by our proposed models outperformed their natural counterparts encoding the same proteins on computational metrics for stability and expression. 
They also demonstrated enhanced performance against the benchmark ORFs used in mRNA vaccines for the SARS-CoV-2 viral spike protein and the varicella-zoster virus (VZV). 
These results highlight the potential of adapting PPLM for designing ORFs tailored to encode target antigens in mRNA vaccines.

% \paragraph*{\textbf{Results}}

% \textbf{Methods:} The abstract serves both as a general introduction to the topic and as a brief, non-technical summary of the main results and their implications. The abstract must not include subheadings (unless expressly permitted in the journal's Instructions to Authors), equations or citations. As a guide the abstract should not exceed 200 words. Most journals do not set a hard limit however authors are advised to check the author instructions for the journal they are submitting to.

% \textbf{Results:} The abstract serves both as a general introduction to the topic and as a brief, non-technical summary of the main results and their implications. The abstract must not include subheadings (unless expressly permitted in the journal's Instructions to Authors), equations or citations. As a guide the abstract should not exceed 200 words. Most journals do not set a hard limit however authors are advised to check the author instructions for the journal they are submitting to.

% \textbf{Conclusion:} The abstract serves both as a general introduction to the topic and as a brief, non-technical summary of the main results and their implications. The abstract must not include subheadings (unless expressly permitted in the journal's Instructions to Authors), equations or citations. As a guide the abstract should not exceed 200 words. Most journals do not set a hard limit however authors are advised to check the author instructions for the journal they are submitting to.
}

\keywords{Language Model, Codon Optimization}



\maketitle

%%%%%%%%%%%%%%%%%%%%%%%%%%%%%%%%%%%%%%%%%%%%%%%%%%%%%%%%%%%%%%%%%%
% Adding extra sections can be done by using the below commands.
% \section{This is an example for first level head---section head}\label{sec3}
% \subsection{This is an example for second level head---subsection head}\label{subsec2}
% \subsubsection{This is an example for third level head---subsubsection head}\label{subsubsec2}
% Sample body text. Sample body text. 
%%%%%%%%%%%%%%%%%%%%%%%%%%%%%%%%%%%%%%%%%%%%%%%%%%%%%%%%%%%%%%%%%%%
% 

\section*{Main}\label{sec1}
mRNA vaccines have demonstrated effectiveness, safety, and scalability in limiting the spread of COVID-19 (SARS-CoV-2)~\cite{vitiello2021brief}. 
However, achieving high expression and stability in mRNA design remains a significant challenge. 
Low expression and stability can lead to reduced immunogenicity, compromising the vaccine's effectiveness. 
An mRNA sequence used in vaccines comprises several components, including the CAP region, untranslated regions (UTRs), and the open reading frame (ORF). 
The ORF, also known as the coding region, is critical for applications such as mRNA vaccines, as it encodes the target antigen protein~\cite{kim2024computational}. 
Therefore, optimizing ORF sequences is an important research problem in therapeutics.

The ORF comprises non-overlapping triplets of nucleotides (`A', `U', `G', and `C'), known as codons, where each codon encodes a specific amino acid in the protein sequence.
There are a total of $4^{3} = 64$ codons for $20$ naturally occurring amino acids. 
Multiple codons can encode the same amino acid; these are referred to as synonymous codons which is the cause for degeneracy in genetic code.
This leads to an exponentially large candidate space of ORFs encoding the same target protein.
However, the preference for one synonymous codon over another, known as codon usage bias, plays a pivotal role in influencing gene expression and stability within a host organism. 
A critical determinant of codon usage bias is the availability of transfer RNA (tRNA) molecules, which vary across organisms for specific codons.
Codons corresponding to abundant tRNAs are translated more efficiently, minimizing ribosomal stalling.
Reduced ribosomal stalling enhances the production of the target antigen protein, which is directly linked to increased immunogenicity~\cite{earle2021evidence, calvo2009upstream}.
As a result, codon optimization focuses on selecting the most appropriate codons among synonymous options to enhance expression within a host species.
Thus, ORF codon optimization must consider both expression and stability to ensure therapeutic efficacy~\cite{kim2022modifications, kudla2006high}.

Codon optimization methods have been employed in the past, yielding improvement in ORF expression and stability.
Traditional codon optimization approaches commonly involve replacing rare codons with more frequently used codons based on the natural codon distribution of the host organism~\cite{welch2009you}.
Codon optimization tools follow similar methodologies~\cite{ranaghan2021assessing}.
However, directly replacing rare codons with the most frequent ones is not always ideal.
This approach can lead to tRNA imbalance, which increases the risk of ribosomal stalling and abrupt termination of translation~\cite{nedialkova2015optimization}.
In earlier studies, stability has often been quantified using the minimum free energy (MFE) of the RNA secondary structure~\cite{zhang2023algorithm}. 
Additionally, the expression level of the optimized sequence is typically evaluated using the codon adaptation index (CAI)~\cite{sharp1987codon}, which measures how closely a sequence's codon usage aligns with the host organism's preferred codons.

Recent advancements in deep learning have been pivotal in various domains of bio-informatics, specifically in genomics.
Given the sequential nature of genomic data, recurrent neural networks (RNNs) and transformer-based architectures~\cite{fallahpour2024codontransformer, constant2023deep, ren2024codonbert} have demonstrated the ability to capture rich intrinsic properties of biological sequences.
This makes them well-aligned and effective for tasks like codon optimization.
Codon optimization has been reformulated in prior studies as either a sequence-tagging task~\cite{goulet2023codon, jain2023icor, gong2023integrated, fu2020codon} or a machine translation task~\cite{ren2024codonbert}. 
Both approaches take the target protein sequence as input and represent each amino acid using context-aware embeddings generated by bidirectional encoder models, such as bidirectional long short-term memory (Bi-LSTM) networks~\cite{fu2020codon} or bidirectional transformer encoder networks~\cite{ren2024codonbert, fallahpour2024codontransformer, yang2019generative}.
These deep learning models leverage training datasets comprising protein sequence and ORF sequence pairs which are specific to a given species to learn context-rich amino acid representations. 
The models then map these representations to optimal codons, achieving enhanced expression in codon optimization tasks.

In this work, we approach codon optimization as a sequence-tagging task, similar to the method adopted in Fu et al.~\cite{fu2020codon}.
However, unlike prior studies, we leverage the rich representations of amino acids extracted from a pre-trained protein language model (PPLM), ProtBert~\cite{elnaggar2021prottrans}, instead of training the model for them.
By doing so, we reformulate codon optimization as a simpler transfer learning task, where only the layers required for mapping amino acids to their optimal codons are fine-tuned. 
This helps to eliminate the heavy task of training the model weights for learning amino acid representations.
To restrict the label space of codons during the tagging task, we pass it through a codon mask that suppresses non-synonymous codons.
Our approach produces species-specific models that can be retrained efficiently for other organisms as needed.
We fine-tune models for species, specifically humans, Escherichia coli (E.coli), and Chinese Hamster Ovary (CHO) cells.
We evaluated the optimized ORF generated by our model, trained on human genes, for the SARS-CoV-2 virus (Wuhan variant spike protein). 
The optimized sequence demonstrated superior performance compared to industry-standard vaccines, Pfizer and Moderna~\cite{teo2022review}, based on computational metrics such as the codon adaptation index (CAI) and minimum free energy (MFE). 
Similarly, we observed significant improvements in the optimized ORF of the VZV encoding gE protein responsible for shingles.
Overall, this work presents a simplified yet effective method for codon optimization that leverages the capabilities of PPLMs.
Our approach could be used as a tool for designing the coding regions encoding target viral antigens useful for mRNA vaccines.

\section*{Results}
\subsection*{Codon optimization using PPLM}
Deep learning models are proficient in capturing underlying data distribution~\cite{lecun2015deep}.
The codon choice among synonymous codons for encoding an amino acid is not random and follows certain selection rules respective to host species~\cite{parvathy2022codon}.
% In this work to embed the codon distribution of human Hg19 genes, codon optimization task reformulation as a sequence tagging task is adapted from the work of Fu, et.al~\cite{fu2020codon}
In this work, the codon optimization task is reformulated as a sequence tagging task to embed the codon distribution of human Hg19 genes, following the approach of Fu et al.~\cite{fu2020codon}.
In this approach, each amino acid in the input protein sequence is treated analogously to a word (or individual token) in a natural language task. 
At each time step, the objective is to predict the optimal codon from 61 possible codons, making the sequence tagging for each amino acid a multi-label classification task.
In this paper, we utilize ProtBert's amino acid representations as embeddings to fine-tune the adapter and selective modules during the fine-tuning phase.
The one in which we only fine-tune the adapter module on top of ProtBert is referred to as Adaptive-ProtBert, whereas the one in which we fine-tune the last layer of ProtBert along with the adapter module is referred to as Adasel-ProtBert.
Since not all 61 possible codons but only a few of them code for an amino acid, previous methods like `codon-box'~\cite{fu2020codon} have been utilized to reduce the label space. 
On a similar premise, we reduce the label space of the sequence tagging task by introducing a codon mask which we term as the `valid-codon' method.

\subsection*{Valid-Codon}
The final layer of the model performs the multi-label classification task by predicting probability scores of 61 codons in a time-distributed manner i.e. at each amino acid. 
To reduce and restrict the model's learning to only synonymous codons we introduce 
a codon mask at the final layer as described in Eq.~\ref{eq:codon-masking}.
Codon mask is a vector $CV \in \mathbb{R}^{61\times 1}$, where non-synonymous codons for a given amino acid are assigned a high negative value of $-10^{9}$.
The tensor $output\_logits_{t} \in \mathbb{R}^{61\times 1}$, has raw probability scores for each of the 61 codons at a given amino acid. 
        \begin{equation}
            masked\_logits = \sum_{t=0}^{L}output\_logits_{t} + CV_{t^{th} amino\_acid}
            \label{eq:codon-masking}
        \end{equation}

\begin{figure}[htbp]
 % Caption and label go in the first argument and the figure contents
 % go in the second argument
{
\centering
  \label{fig:codon optimization as sequence tagging}
  {\includegraphics[width=0.9\linewidth]{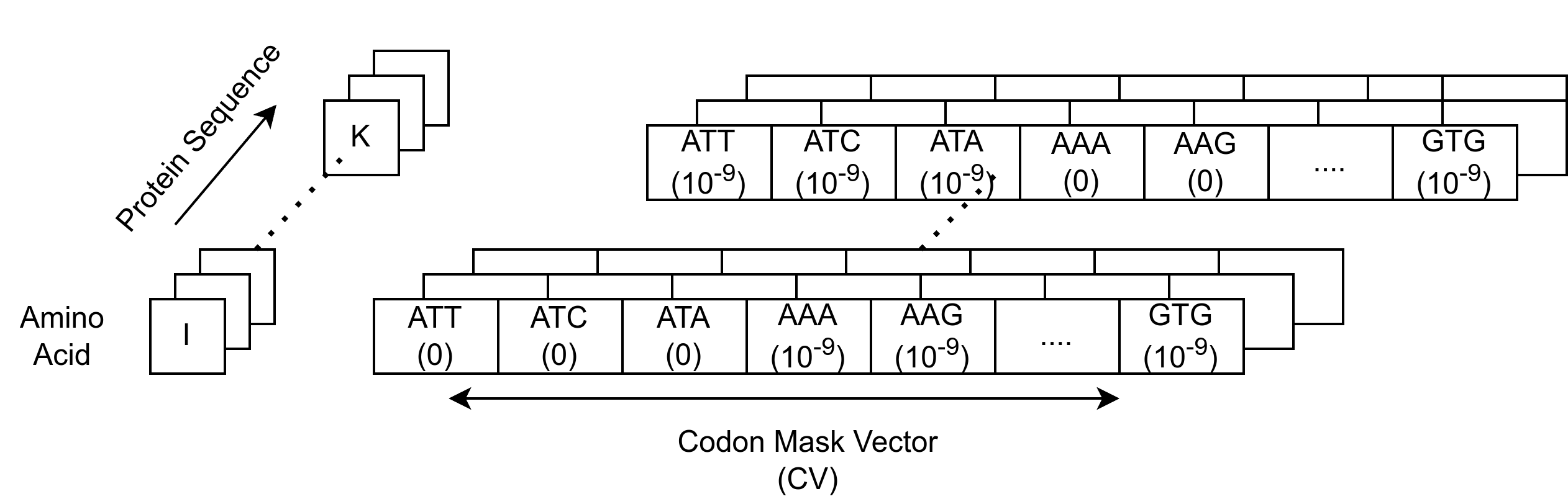}}}
  \caption{Codon Mask for each $t^{th}$ amino acid ($CV_{t^{th} amino\_acid}$)}
\end{figure}
E.g., in the case of Alanine (A), out of 61 codon classes the ones that do not correspond to `GCT', `GCC', `GCA', and `GCG' are assigned a high negative value in $CV$ mask for amino acid A.

\subsection*{Experimental Results}
\subsubsection*{Species generalization}
In this study, three models were trained by fine-tuning the PPLM on protein sequences specific to a particular species: humans (Hg19 genes), E.coli, and Chinese Hamster Ovary (CHO) cells. 
These models are referred to as Adasel-ProtBert-short, Adasel-ProtBert-E.coli, and Adasel-ProtBert-CHO, respectively.
To ensure that each model is optimized for its target species and to evaluate its generalizability, the optimized ORFs from each fine-tuned model were cross-validated on test sets for the rest of the other species. 
This cross-species validation demonstrates the adaptability of the fine-tuned models and highlights the species-specific nature of codon optimization, while also showcasing their potential for generalization to different host species.

\begin{figure}[!htbp]
    \centering
    \subfigure[][Human as a host species for training]{\label{fig:Res-Hg19-fine-tuned}%
        \includegraphics[width=0.9\linewidth]{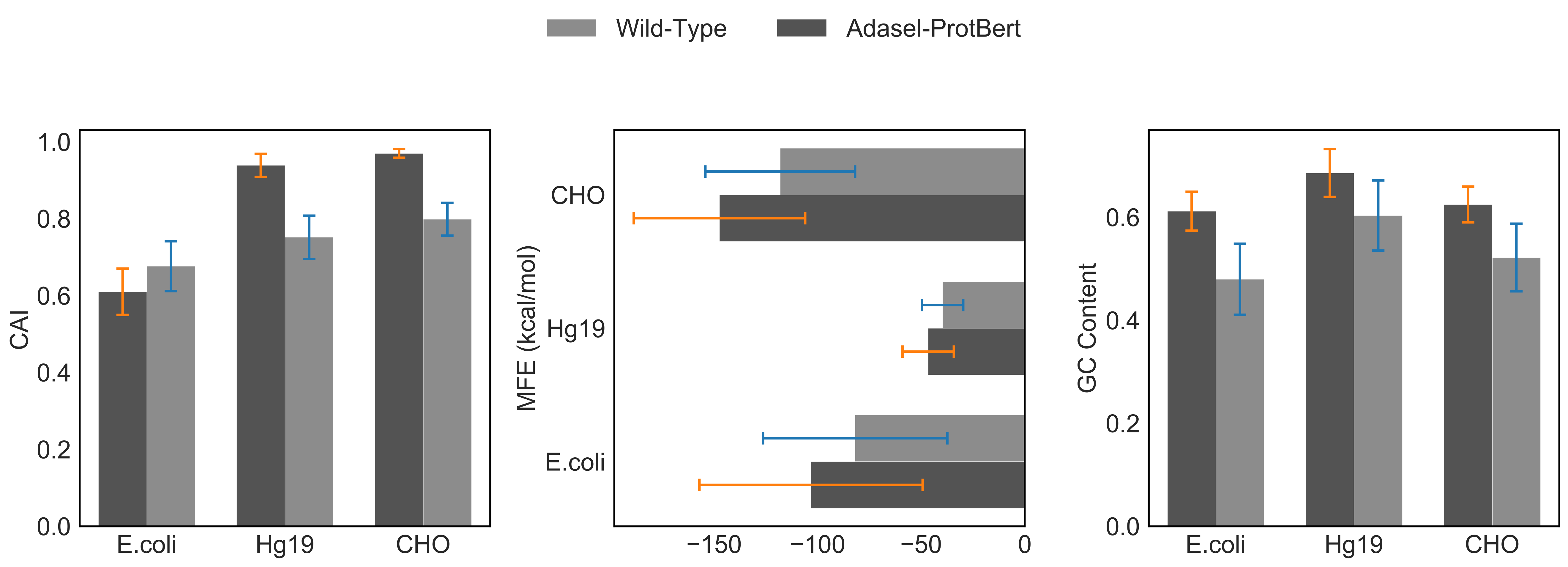}}%
      \vspace{0.3cm}
      
    \subfigure[][E.coli as a host species for training]{\label{fig:Res-E.coli-fine-tuned}%
        \includegraphics[width=0.9\linewidth]{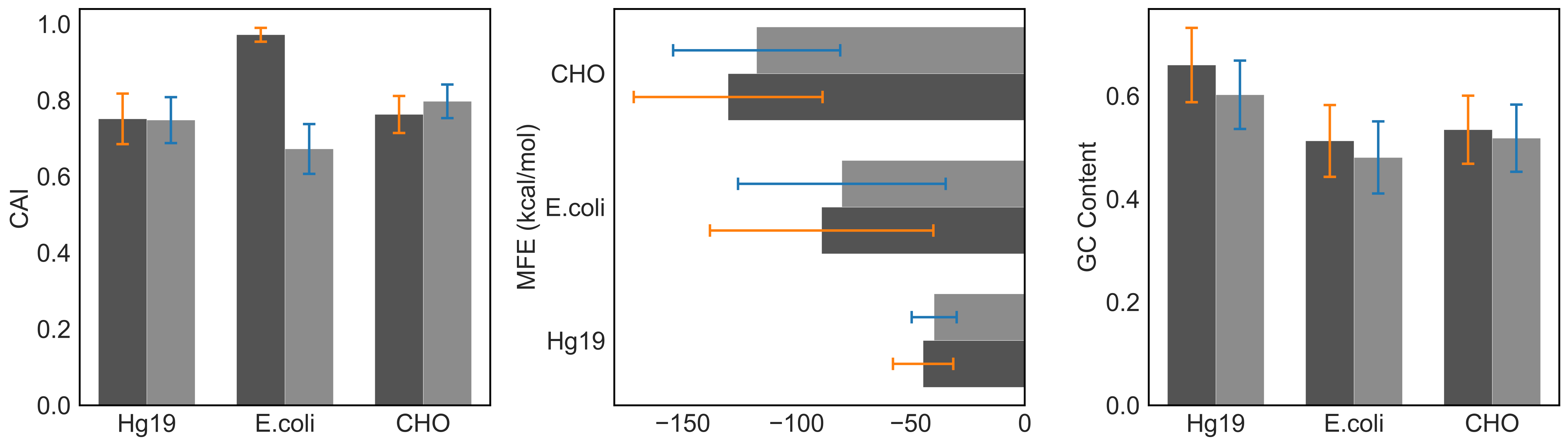}}
        
    \vspace{0.3cm} % Adds vertical space between rows

    \subfigure[][CHO as a host species for training]{\label{fig:Res-CHO-fine-tuned}%
    \includegraphics[width=0.9\linewidth]{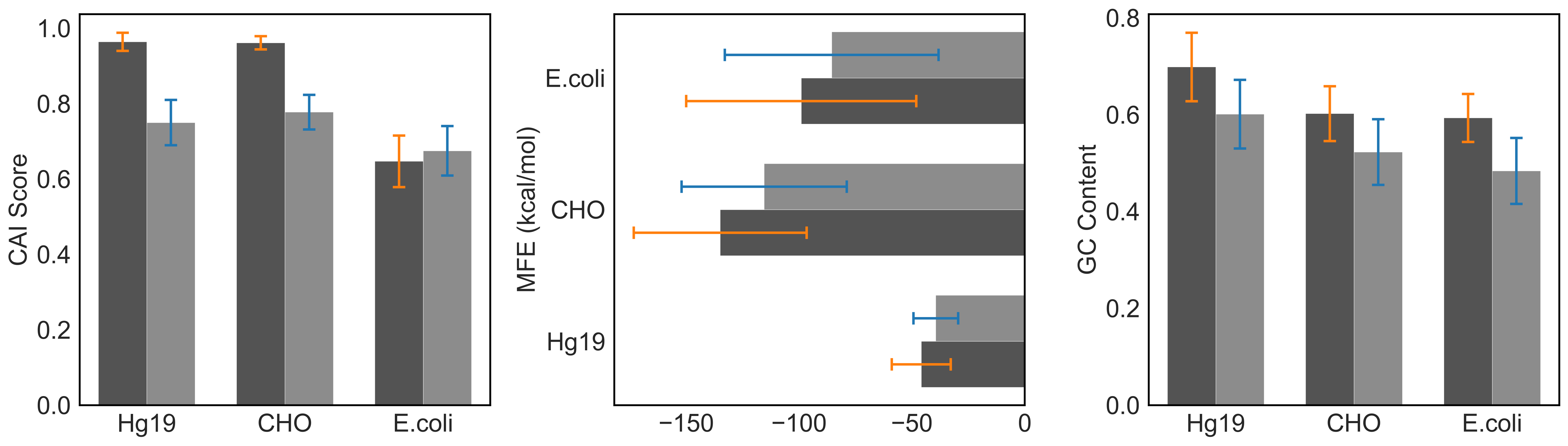}}
    \caption{In this figure we compare the results of different species (Human, E.coli and CHO) on their fine-tuned model ORFs from Adasel-ProtBert-short, Adasel-ProtBert-E.coli and Adasel-ProtBert-CHO respectively.
    (a) The expression and stability evaluation of Adasel-ProtBert-short on ORF optimization across species with their respective wild type ORFs.
    (b) The expression and stability evaluation of Adasel-ProtBert-E.coli on ORF optimization across species with their respective wild type ORFs.
    (c) The expression and stability evaluation of Adasel-ProtBert-CHO on ORF optimization across species with their respective wild type ORFs.}
    \label{fig: Species specific codon optimization}
\end{figure}

On fine-tuning with humans as the host species, the Adasel-ProtBert-short model produced optimized ORFs that showed substantial improvements in expression metrics compared to their baseline human wild-type ORFs (p $<$ 0.001).
However, when validated on E.coli\_dataset and CHO\_dataset protein sequences for codon optimization, the results revealed an expected decrease in the CAI metric compared to their respective wild-type ORFs, except for CHO (Fig.~\ref{fig:Res-Hg19-fine-tuned}). 
This exception is attributed to the similarity found in codon distribution between CHO and Hg19.
Similarly, the Adasel-ProtBert-E.coli model showed improved performance against the wild-type ORFs in the E.coli\_dataset test set, achieving improved CAI metrics (p $<$ 0.001). 
However, its performance dropped significantly when evaluated on human and CHO protein sequences for expression  (Fig.~\ref{fig:Res-E.coli-fine-tuned}).
For the Adasel-ProtBert-CHO model, comparable trends were observed, with improved CAI values for CHO against its Wild Type ORFs (p $<$ 0.001) but reduced performance for E.coli(Fig.~\ref{fig:Res-CHO-fine-tuned}).

We conducted an additional experiment by extending our codon optimization approach to longer sequences, using the Hg19\_long\_filtered\_mfe dataset with humans as the host species. The Adasel-ProtBert-long-mfe model demonstrated improved performance, achieving a CAI of 0.97 on the Hg19\_long\_filtered\_mfe test set.
To evaluate the impact of MFE filtering, we trained the Adasel-ProtBert model on the Hg19\_random\_mfe dataset, referred to as Adasel-ProtBert-random-mfe in this study. Without MFE filtering, there was a noticeable decline in both CAI and MFE metrics on the test set. The average CAI dropped by 10.3\% to 0.87, while the MFE increased substantially by 75\%, leading to reduced expression and stability of the optimized ORFs.

\subsubsection*{Result for mRNA vaccine design}
To evaluate the broader applicability of our approach in mRNA vaccine design, we optimized ORFs encoding the SARS-CoV-2 spike protein and the VZV gE protein. 
The evaluation was based on three critical computational metrics: CAI, MFE, and GC-Content. 
The CAI metric, which reflects the expression of ORF, was computed using Hg19 genes as the reference sequence set. 
These metrics provided a comprehensive view of the expression efficiency and structural stability of the optimized ORFs, critical factors for successful mRNA vaccine development.

\begin{figure}[htbp]
          \centering
          \subfigure[][CAI and GC Content Comparison]{\label{fig:CAI-GC-COVID}%
            \includegraphics[width=0.5\linewidth]{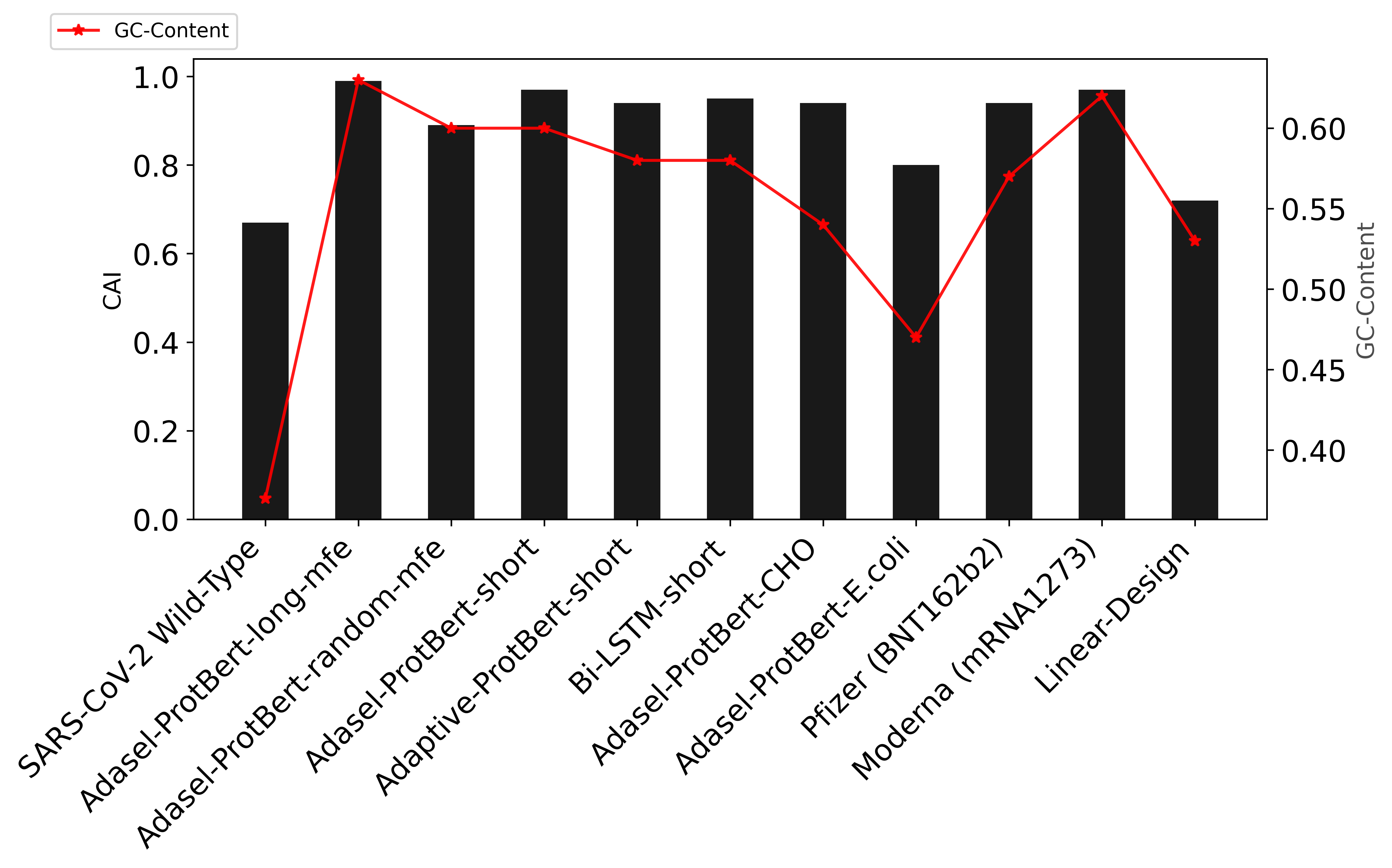}}%
          \hfill
          \subfigure[][MFE Comparison]{\label{fig:MFE-COVID}%
            \includegraphics[width=0.5\linewidth]{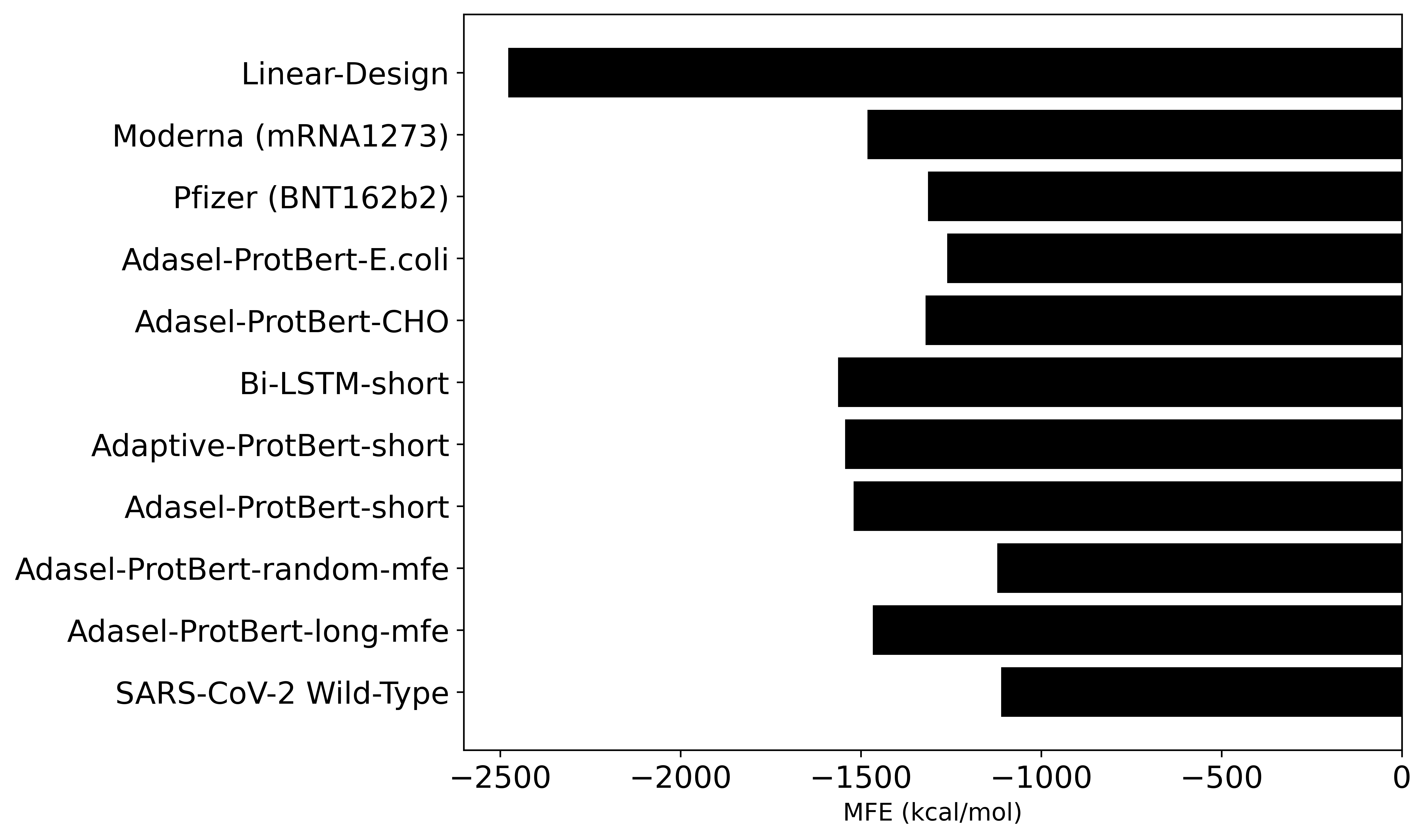}}
            
          \vspace{0.5cm} % Adds vertical space between rows
    
        \subfigure[][CAI vs. MFE]{\label{fig:CAI_MFE_COVID-19}%
        \includegraphics[width=0.7\linewidth]{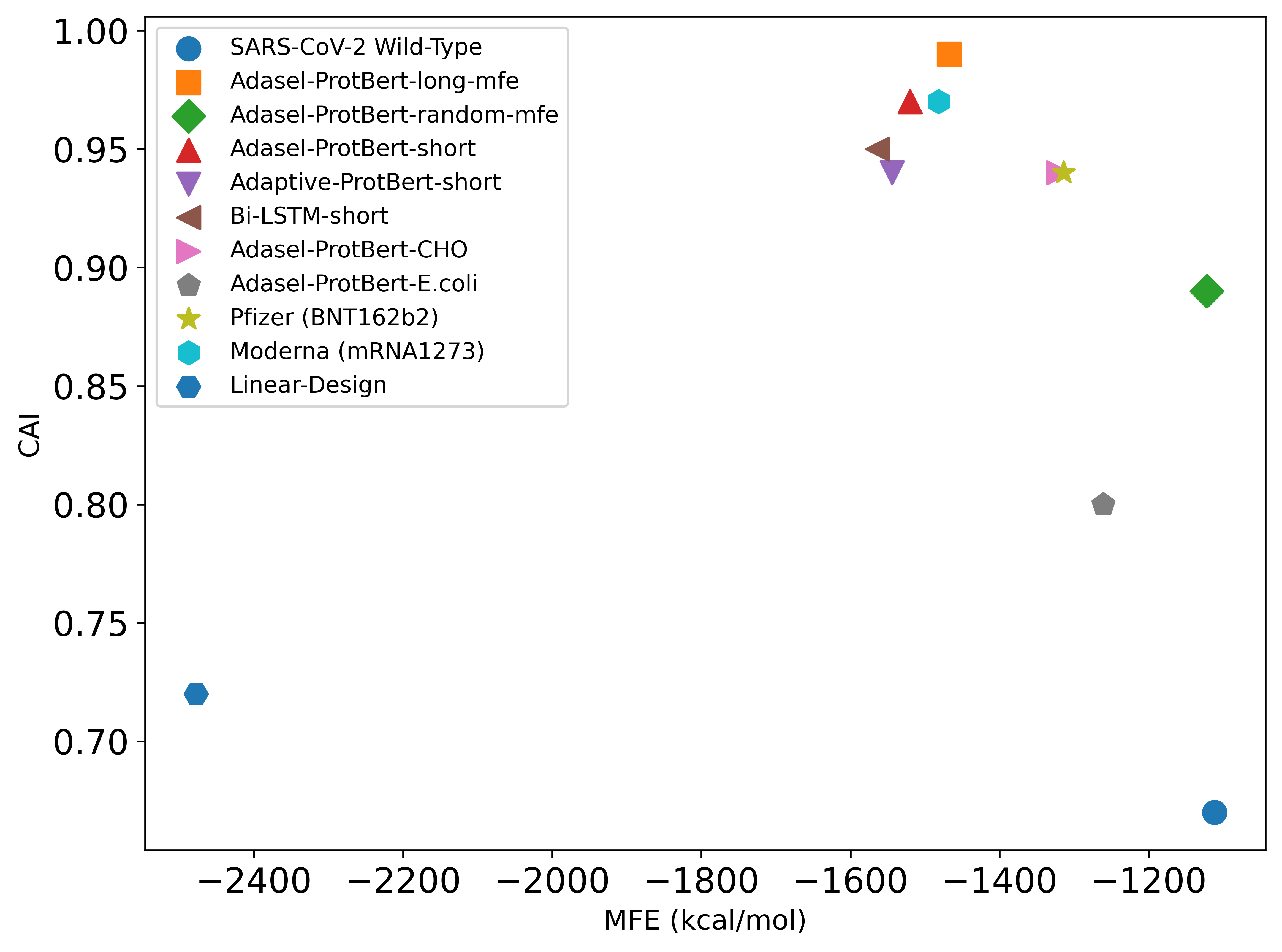}}
        \caption{ORFs from Adasel-ProtBert-long-mfe, Adasel-ProtBert-random-mfe, Adasel-ProtBert-short, Adaptive-ProtBert-short, and Bi-LSTM are the ones that were trained on Hg19 genes, whereas Adasel-ProtBert-Ecoli and Adasel-ProtBert-CHO are the models trained on E.coli and CHO genes respectively. The Covid19-Wild-Type is the naturally occurring ORF sequence found in the SARS-CoV-2 virus.
        (a) Th ORF sequences from design types with their CAI (the higher the better) and GC-Content (optimal range is 30-70\%) (b) Different design types structural stability values in terms of MFE (lower the MFE value, the higher the stability). (c) Comparison of expression level with stability for ORFs are illustrated here}
        \label{fig: SARS-COVID-19 results}
    \end{figure}

The benchmark ORFs encoding the SARS-CoV-2 spike protein were sourced from Pfizer's BNT162b2, Moderna's mRNA-1273, and the widely used codon optimization tool Linear-Design~\cite{zhang2023algorithm}. 
The naturally occurring wild-type sequence, referred to as SARS-CoV-2 Wild Type, served as the baseline.
SARS-CoV-2 Wild Type had a CAI of 0.67, reflecting suboptimal adaptation to humans.
Adasel-ProtBert-long-mfe optimized ORF achieved the highest CAI of 0.99 (Fig.\ref{fig:CAI-GC-COVID}) with MFE of -1467.3 kcal/mol(Fig.~\ref{fig:MFE-COVID}), superior to benchmark and baseline ORFs.
The second best performing model was Adasel-ProtBert-short which exhibits slightly lower CAI than the best one but had a more stable sequence than it(Fig.~\ref{fig:CAI_MFE_COVID-19}).
Illustrated in Fig.~\ref{fig:CAI_MFE_COVID-19}, the fine-tuned models were better than the industry vaccines on our computational metrics.
For all of our fine-tuned models, the GC-Content was in the optimal range~\cite{kim2024computational}.

\begin{figure}[htbp]
          \centering
          \subfigure[][CAI and GC Content Comparison]{\label{fig:CAI-GC-VZV}%
            \includegraphics[width=0.5\linewidth]{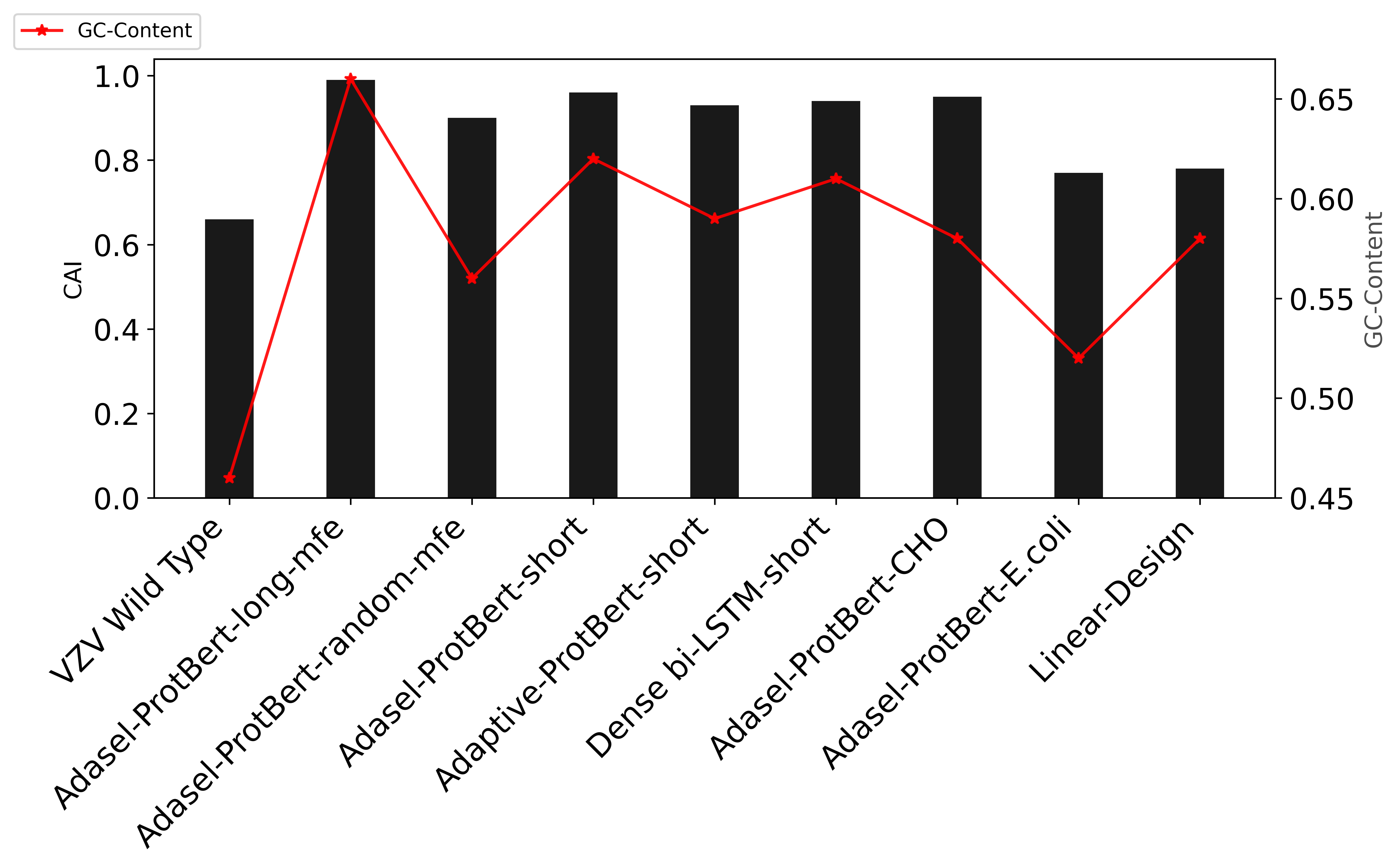}}%
          \hfill
          \subfigure[][MFE Comparison]{\label{fig:MFE-VZV}%
            \includegraphics[width=0.5\linewidth]{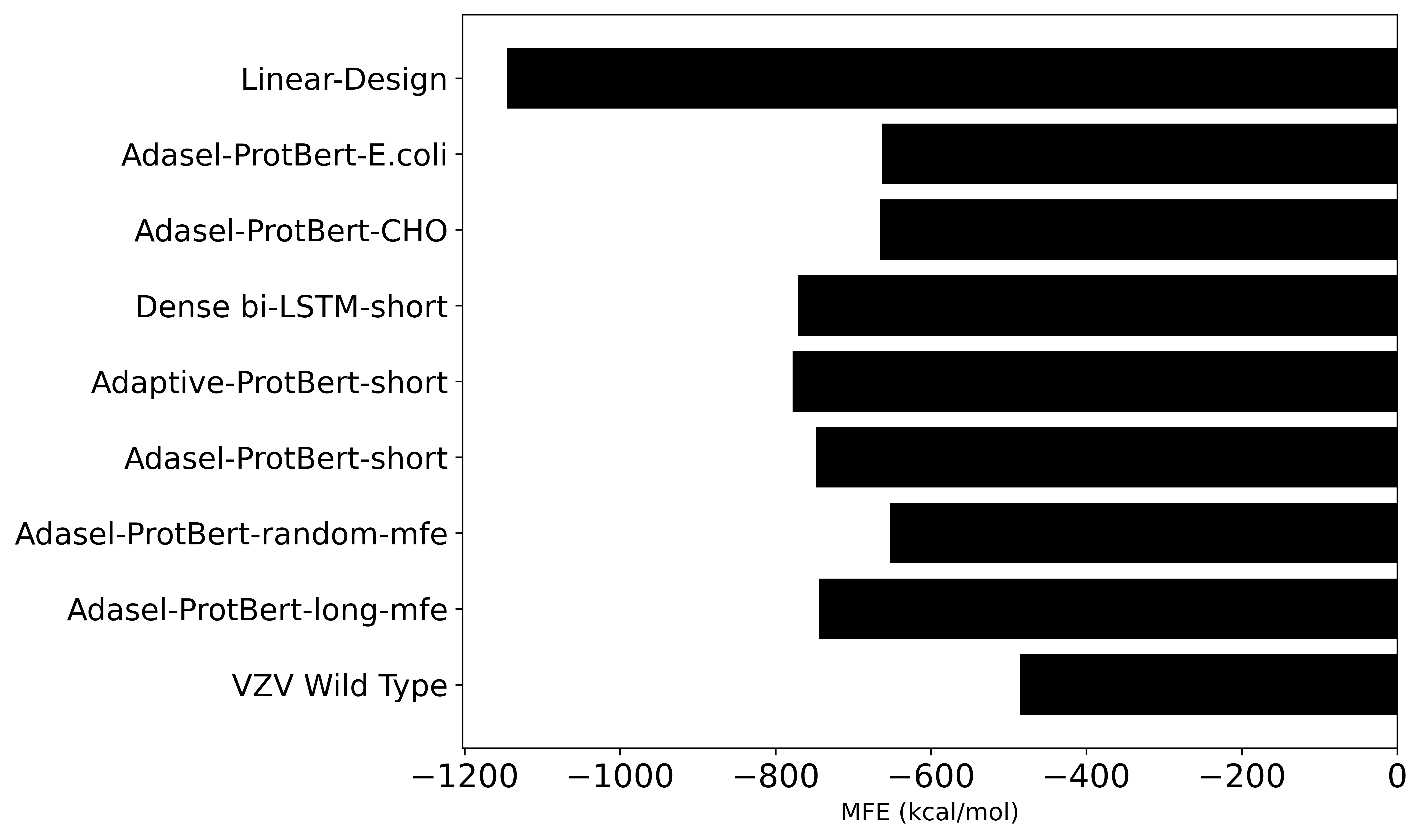}}
            
          \vspace{0.5cm} % Adds vertical space between rows
    
        \subfigure[][CAI vs. MFE]{\label{fig:CAI_MFE_VZV}%
        \includegraphics[width=0.7\linewidth]{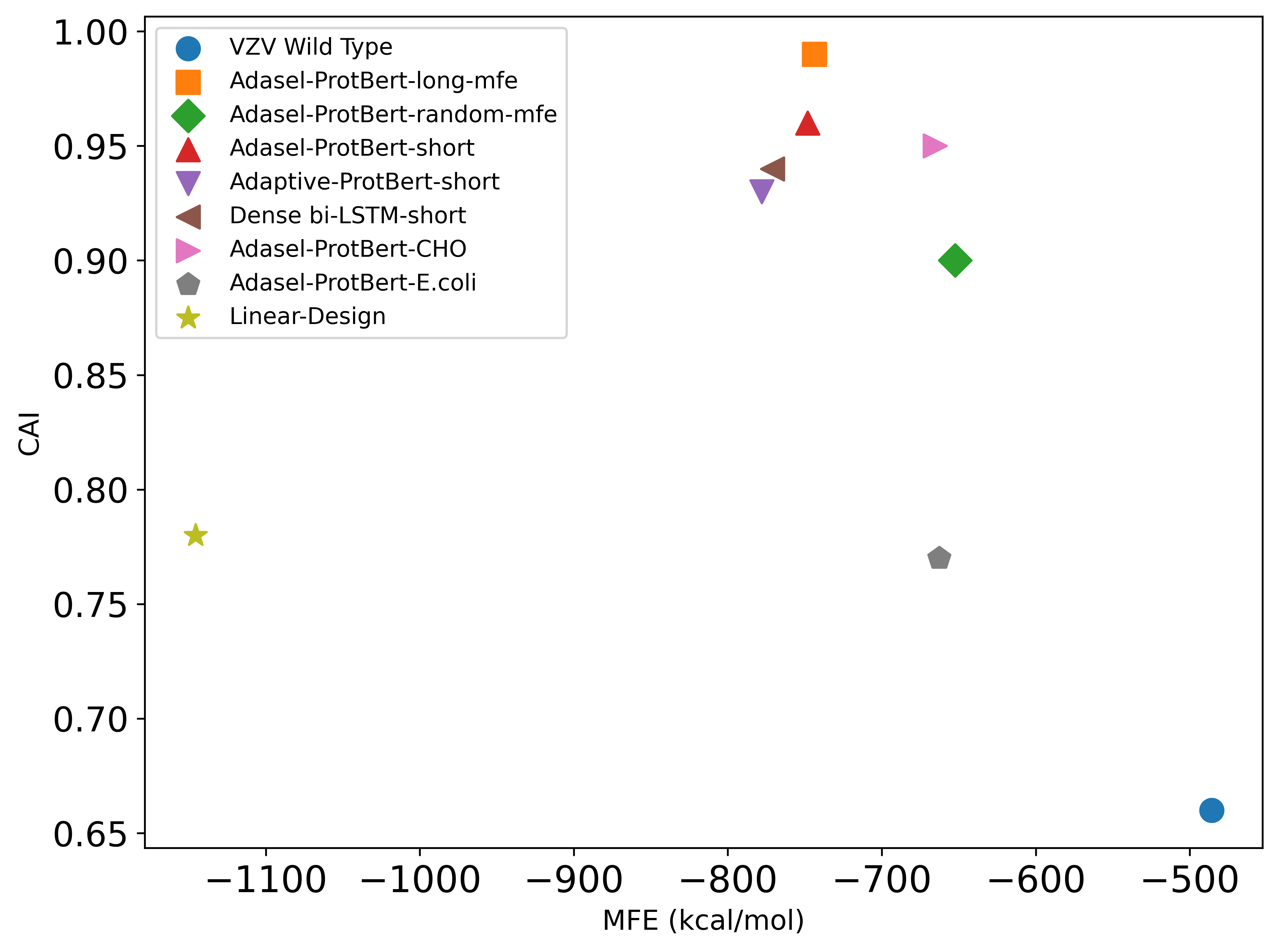}}
        \caption{The design types Adasel-ProtBert, Adaptive-ProtBert, and Bi-LSMT are models trained on Hg19 genes, whereas Adasel-ProtBert-Ecoli and Adasel-ProtBert-CHO are trained on E.coli and Chinese-Hamster genes respectively. Wild-Type is the naturally occurring ORF sequence found in the VZV.
        (a) Th ORF sequences from design types with their CAI (the higher the better) and GC-Content (optimal range is 30-70\%) (b) Different design types structural stability values where lower the MFE value, the higher the stability. (c) Comparison of expression level with stability for ORFs are illustrated here}
        \label{fig: VZV results}
    \end{figure}

For the VZV gE-protein-encoding ORFs, benchmark sequences were derived using the Linear-Design tool, similar to the approach used for the SARS-CoV-2 spike protein. 
As the sequence of Moderna's mRNA vaccine (mRNA-1273) for VZV remains proprietary and unavailable, it could not be included in the analysis. 
The baseline ORF sequence for the VZV gE protein was the naturally occurring wild-type sequence from the virus, referred to as VZV Wild Type.
Similar to the results observed for the SARS-CoV-2 spike protein encoding design of ORFs, our fine-tuned models optimized ORF encoding VZV gE protein were better than the benchmark and baseline ORF sequences in terms of expression (Fig.~\ref{fig:CAI-GC-VZV}) and stability (Fig.~\ref{fig:MFE-VZV}).
The GC-Content for ORFs optimized using our fine-tuned models was balanced and in the appropriate range.
Refer to Table~\ref{tab:res-SARS-CoV-2} and Table~\ref{tab:res-VZV} for quantitative results.

The superior CAI and MFE values of our optimized sequences, combined with appropriately balanced GC-Content, demonstrate their suitability for human translational machinery while maintaining the structural stability required for mRNA vaccine efficacy. 
This dual optimization, validated across two distinct viral targets, underscores the versatility and robustness of our approach in mRNA vaccine design.

    \begin{table}[!htbp]
        \centering
            \caption{Quantitative analysis of ORFs optimized by our model with benchmark and baseline ORFs on computational metrics-CAI, MFE, and GC-Content for the COVID-19 vaccine}
            \begin{tabular}{llcc}
                \hline
                \textbf{\textbf{ORF Sequence}}  & \textbf{\textbf{CAI}} & \textbf{\textbf{MFE (kcal/mol)}} & \textbf{\textbf{GC-Content}} \\ \hline
                
                \multicolumn{1}{|l}{SARS-CoV-2 Wild Type} &  0.67                 & -1111.5                          & \multicolumn{1}{c|}{0.37}    \\ \hline
                Adasel-ProtBert-long-mfe       &  0.99                 & -1467.3                          & 0.63                         \\
                Adasel-ProtBert-random-mfe     &  0.89                 &  -1122.0                         & 0.60       \\
                Adasel-ProtBert-short          &  0.97                 & -1520.1                          & 0.60       \\
                Adaptive-ProtBert-short        & 0.94                  & -1544.0          & 0.58                \\
                Bi-LSTM-short            & 0.95        & -1563.7          & 0.58                        \\
                Adasel-ProtBert-CHO             & 0.94 & -1321.0                   & 0.54        \\
                Adasel-ProtBert-E.coli          & 0.80  & -1260.9                  & 0.47       \\
                Pfizer (BNT162b2)              & 0.94                  & -1314.1                         & 0.57                        \\
                Moderna (mRNA1273)             & 0.97 & -1481.8                         & 0.62        \\
                Linear-Design                  & 0.72                 & -2477.7          & 0.53                       \\ \hline
            \end{tabular}
            \label{tab:res-SARS-CoV-2}
    \end{table}

% \subsection*{Results on VZV virus mRNA vaccine}
    
    \begin{table}[!htbp]
        \centering
            \caption{Quantitative analysis of ORFs optimized by our model with benchmark and baseline ORFs on computational metrics-CAI, MFE, and GC-Content for the VZV virus vaccine}
            \begin{tabular}{llcc}
                \hline
                \textbf{\textbf{ORF Sequence}}  & \textbf{\textbf{CAI}} & \textbf{\textbf{MFE (kcal/mol)}} & \textbf{\textbf{GC-Content}} \\ \hline
                
                \multicolumn{1}{|l}{VZV Wild Type}    &  0.66                 & -485.7                          & \multicolumn{1}{c|}{0.46}    \\ \hline
                Adasel-ProtBert-long-mfe       &  0.99                 & -743.7                          & 0.66                         \\
                Adasel-ProtBert-random-mfe     &  0.90                 & -652.3                          & 0.56       \\
                Adasel-ProtBert-short          &  0.96                 & -748.1                          & 0.62       \\
                Adaptive-ProtBert-short        & 0.93                  & -778.0                          & 0.59                \\
                Bi-LSTM-short            & 0.94                  & -771.0                          & 0.61                        \\
                Adasel-ProtBert-CHO            & 0.95                  & -665.4                          & 0.58        \\
                Adasel-ProtBert-E.coli         & 0.77                  & -662.7                          & 0.52       \\
                % Pfizer (BNT162b2)              & 0.94                  & -1314.1                          & 0.57                        \\
                % Moderna (mRNA1273)             & 0.97                  & -1481.8                          & 0.62        \\
                % Optimum Gene
                Linear-Design                  & 0.78                 & -1145.6                          & 0.58                       \\ \hline
            \end{tabular}
            \label{tab:res-VZV}
    \end{table}
% \textbf{Ask about the ablation study for different length training and  MFE filter and no filter}
% \subsubsection*{Ablation Study}
% \subsubsection*{}

\section*{Discussion}\label{discussion}
An effective computational design of ORF sequences could be of great benefit for finding potential candidate sequences in designing mRNA vaccines.
Designing an optimal ORF that encodes a protein is challenging due to its large search space.
In this work, we present a simpler fine-tuning approach, unlike previous deep-learning approaches for finding optimal ORF through codon optimization.
We evaluated our approach for optimizing ORFs on three different species individually and observed promising results on computational metrics for expression and stability.
Performing codon optimization on different species backs the generalizability of our approach in optimizing ORFs given a host species.
Further, we extended our work to design optimized ORF sequences encoding SARS-CoV-2 spike protein and VZV gE protein for its application in mRNA vaccines for COVID-19 and shingles.
Our optimized ORFs showcased improved metrics computationally against the Pfizer and Moderna industry-approved mRNA vaccines.
We could not compare our ORF for shingles with any benchmark industry-approved vaccines as they are not available in public.
However, our ORF encoding VZV gE protein performed better in terms of CAI against a popular tool Linear-Design.
Overall, our approach could be a potential tool in designing ORF sequences for therapeutic proteins for a given host species.
\section*{Method} 
\subsection*{Data}
ORF sequences and their corresponding proteins were sourced from UCSC and NCBI for the Human (Hg19), E.coli, Chinese-Hamster, Pfizer, and Moderna vaccines.
The benchmark ORF sequences for SARS-CoV-2 spike and VZV gE protein using Linear-Design were taken from the work in ~\cite{zhang2023algorithm}.
The designing of the dataset for training was done by filtering the ORF length and MFE values.
Filtering on MFE values to keep high stable sequences for training ensured joint optimization for codon usage and stability.

For the human species-specific fine-tuning and testing, we curated three datasets- Hg19\_short\_filtered\_mfe, 
Hg19\_long\_filtered\_mfe and Hg19\_random\_mfe (Table.~\ref{tab:datasets})
The E.coli and CHO, training and testing datasets statistics, are mentioned in the Table.~\ref{tab:datasets}

\begin{table}[!htbp]
    \centering
    \caption{Summary of datasets used in this study, including species, ORF maximum lengths, mean MFE values, and filtering status.}
    \resizebox{\textwidth}{!}{%
    \begin{tabular}{llccc}
        \hline
        \textbf{Dataset}               & \textbf{Species} & \textbf{ORF Maximum Length} & \textbf{Mean MFE (kcal/mol)} & \textbf{MFE Filtering} \\ \hline
        Hg19\_short\_filtered\_mfe     & Human            & 150                         & -40.5                        & \checkmark             \\
        Hg19\_long\_filtered\_mfe      & Human            & 2500                        & -224.9                       & \checkmark             \\
        Hg19\_random\_mfe              & Human            & 2500                        & -57.7                        & \texttimes            \\
        E.coli\_dataset                & \textit{E.coli} & 2500                        & -83.3                        & \checkmark             \\
        CHO\_dataset                   & CHO              & 2500                        & -116.9                       & \checkmark             \\ \hline
    \end{tabular}%
    }
    \label{tab:datasets}
\end{table}

% \begin{table}[!htbp]
%     \centering
%     \caption{Caption}
%     \begin{tabular}{ccccc}
%         Dataset & Species & ORF maximum length &  Mean MFE (in kcal/mol) & MFE filtering  \\
%          Hg19\_short\_filtered\_mfe &  Human & 150  &  -40.5 & \checkmark \\
%          Hg19\_long\_filtered\_mfe & Human & 2500 & -224.9 & \checkmark \\
%          Hg19\_random\_mfe & Human & 2500  & -57.7  & \times \\
%          E.coli\_dataset  & E.coli  & 2500 & -83.3  & \checkmark \\
%          CHO\_dataset & CHO & 2500 & -116.9  & \checkmark \\
%     \end{tabular}
%     \label{tab:my_label}
% \end{table}
% Hg19 genes are highly expressive and used as a reference for vaccine development~\cite{gong2023integrated}. 
% To structure the Hg19 genes for train and test data, ORFs were filtered by sequence length and MFE values of ORFs resulting in three distinct datasets.
% Hg19\_short\_filtered\_mfe, Hg19\_long\_filtered\_mfe and Hg19\_random\_mfe[Table~\ref{table}].
% The ORF sequences were filtered on MFE values to ensure training on not only highly expressive Hg19 genes but also on highly stable sequences for structural stability.
% Similar filtering was done for designing training and test sequences for E.coli and CHO datasets.

\subsection*{Model Architecture and Training}
\begin{figure}[!htbp]
 % Caption and label go in the first argument and the figure contents
 % go in the second argument
{
\centering
  {\includegraphics[width=1.0\linewidth]{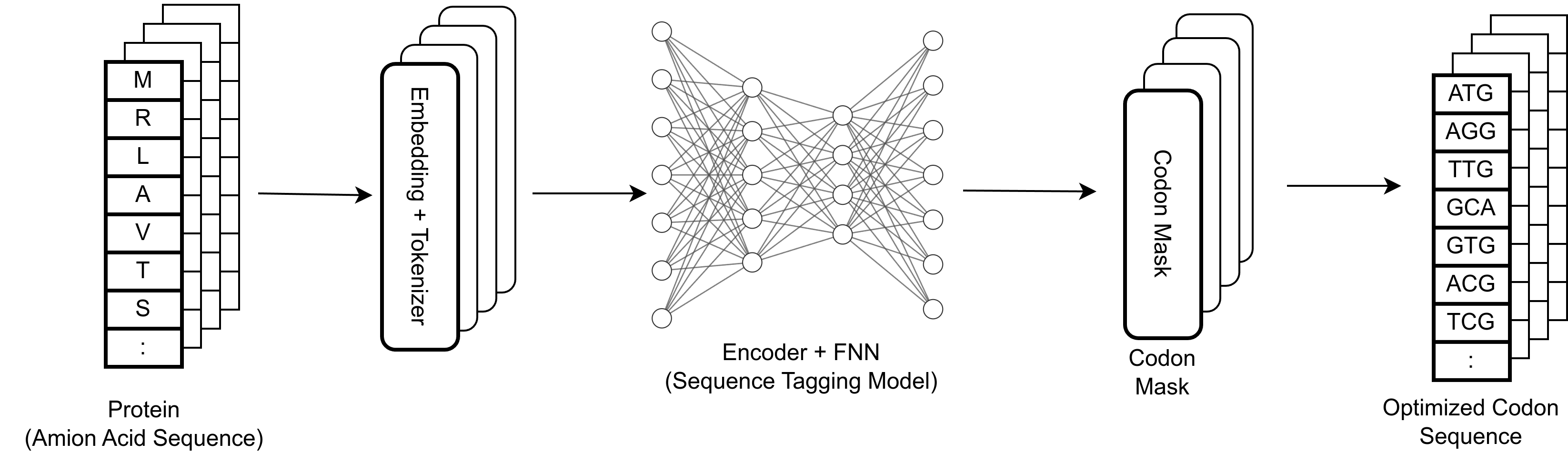}}}
  \label{fig: sequence-tagging-method}
  \caption{Codon Optimization flow chart as a sequence tagging task. First, the input protein sequence is chunked into individual tokens of amino acids. Each tokenized amino acid is passed to a neural network (Encoder) to capture rich context-aware representations. Classifier layer (FNN) + Codon Mask are applied in a time-distributed way to tag optimal codon out of 61 for each amino acid.
  The final output will be a sequence of codons i.e optimized ORF.}
\end{figure}

In this study, codon optimization was framed as a sequence tagging task, in which the input to the model is a protein sequence and the output is a tagged ORF sequence~\cite{goulet2023codon}. 
Each amino acid in the protein sequence is assigned an optimal codon by the model, effectively generating an ORF sequence that maximizes stability and expression.
On the final logits, we apply the `valid-codon` method only during training and it is removed during inference.
Fig.~\ref{fig: sequence-tagging-method} illustrates the method discussed.

\begin{figure}[!htbp]
  \centering
  
  % First subfigure centered on the top row
  \subfigure[]{\label{fig:Protbert-fine-tune}%
  \hspace{-0.5cm}
    \includegraphics[width=0.7\linewidth]{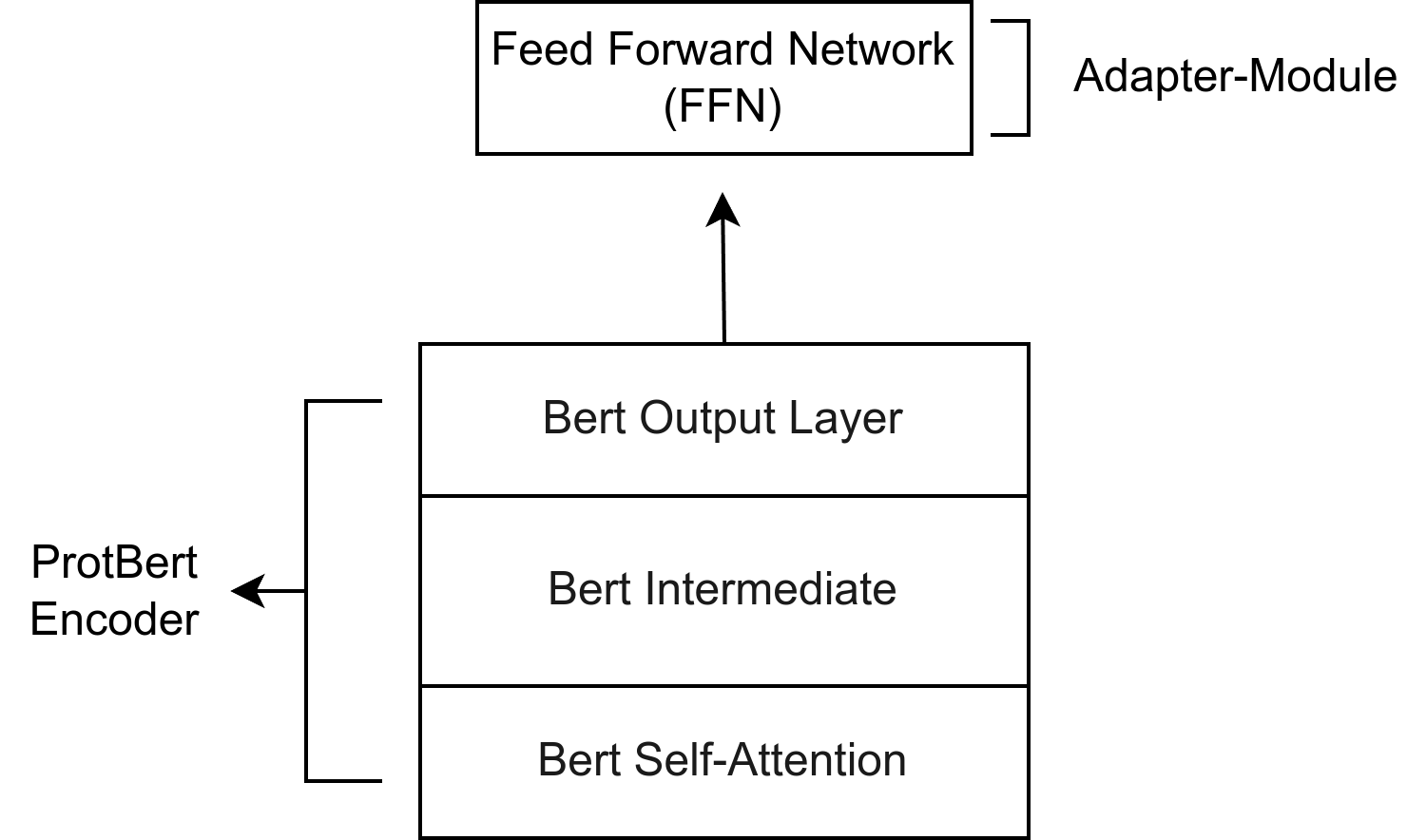}
    }
  
  \vspace{0.5cm} % Adds vertical space between rows
  
  % Second row with two side-by-side subfigures
  \subfigure[]{\label{fig:Adaptive-Protbert}%
    \includegraphics[width=0.3\linewidth]{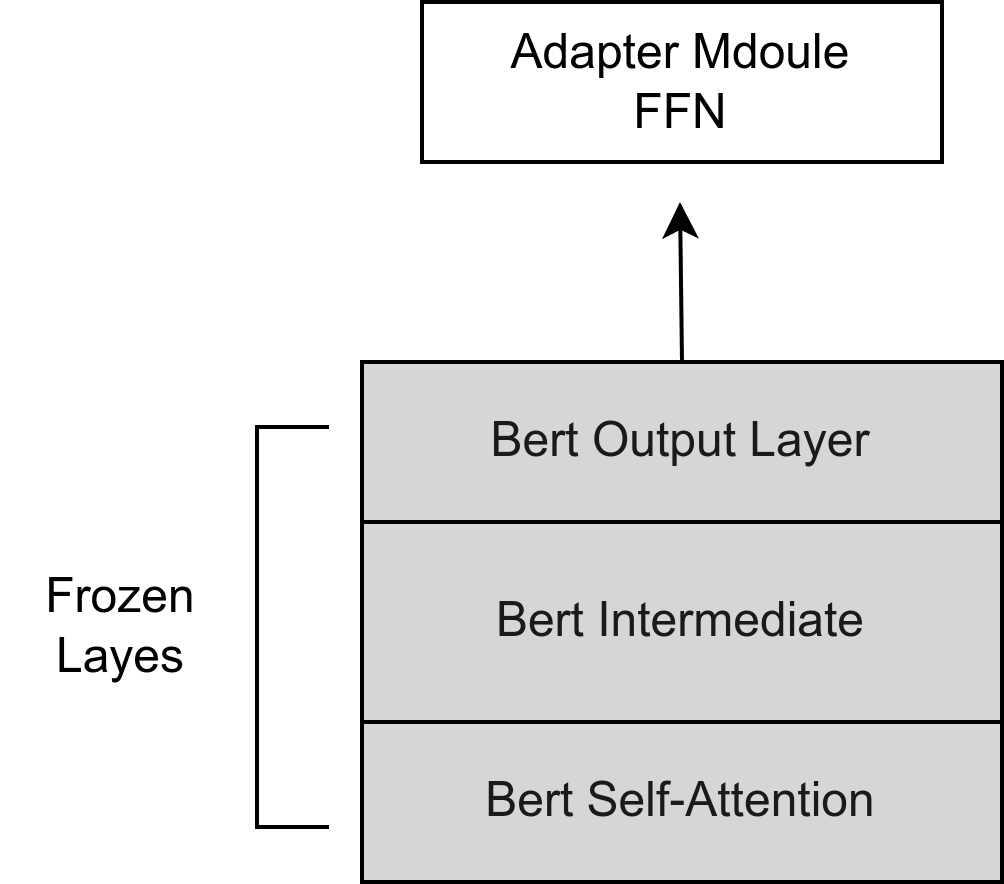}}%
  \hspace{1.5cm}  % Horizontal space between the two bottom figures
  \subfigure[]{\label{fig:Adasel-ProtBert}%
    \includegraphics[width=0.3\linewidth]{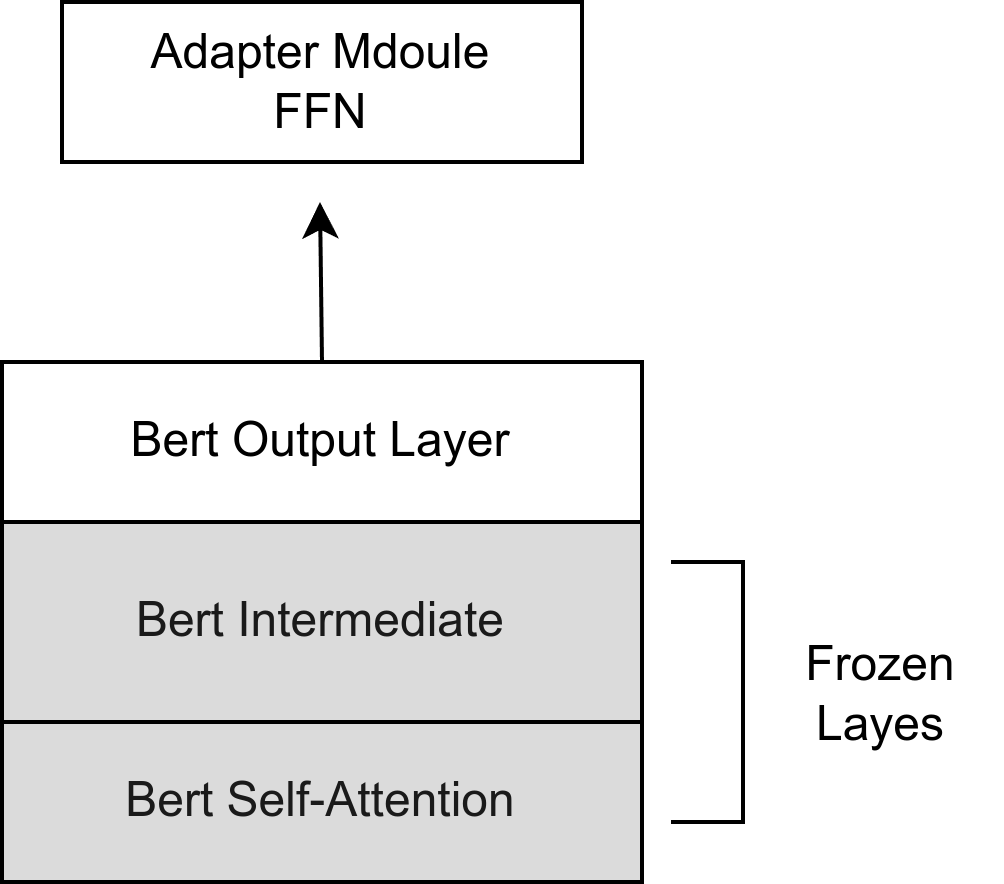}}

  \label{fig:Protbert}
  \caption{In this work, we perform two types of fine-tuning over the ProtBert architecture. Different fine-tuning layers are illustrated.
            (a) ProtBert encoder layer components and the adapter module added over it for classification task is illustrated
            (b) Adaptive-ProtBert architecture where the complete encoder layers are frozen and only the weights of the Adapter Module are trained.
            (c) Adasel-ProtBert architecture where the last layer of the Encoder is unfreezed. The Bert Ouput Layer and Adapter Module weights are trained.}   
\end{figure}
The PPLM~\cite{elnaggar2021prottrans}, ProtBert, originally was trained on 216 million protein sequences to learn their intrinsic chemical and structural properties.
For our downstream task of codon optimization we leveraged their rich amino acid representation as an embeddings for our input protiein sequences.
Fine-tuning ProtBert for codon optimization as an tagging task allowed the model to learn relationships between protein sequences and optimal codon usage patterns, informed by the codon preferences within a specific species.
The fine-tuning was carried out in two ways: adaptive fine-tuning(Adaptive-ProtBert) and adaptive + selective fine-tuning(Adasel-ProtBert), illustrated in Fig.~\ref{fig:Adaptive-Protbert} and Fig.~\ref{fig:Adasel-ProtBert} respectively.

\subsection*{Evaluation Metircs}
The ORFs were evaluated using three primary computational metrics: CAI, Minimum Free Energy (MFE), and GC-Content. 
CAI provided a measure of the codon usage preference relative to the host organism, reflecting potential expression efficiency. 
MFE, calculated using the RNAFold tool~\cite{lorenz2011viennarna} in this work, provides insight into the stability of the predicted ORF sequences, with more negative values indicating greater structural stability. 
GC-Content determines the ratio of 'G' and 'C' nucleotides in the ORF sequence and a range of 0.3 to 0.7 is considered to be optimal.

\backmatter

% \bmhead{Supplementary information}

% If your manuscript includes potentially identifying patient/participant information, or if it describes human transplantation research, or if it reports results of a clinical trial then  additional information will be required. Please visit (\url{https://www.nature.com/nature-research/editorial-policies}) for Nature Portfolio journals, (\url{https://www.springer.com/gp/authors-editors/journal-author/journal-author-helpdesk/publishing-ethics/14214}) for Springer Nature journals, or (\url{https://www.biomedcentral.com/getpublished/editorial-policies\#ethics+and+consent}) for BMC.

% \section{Discussion}\label{sec12}

% Discussions should be brief and focused. In some disciplines use of Discussion or `Conclusion' is interchangeable. It is not mandatory to use both. Some journals prefer a section `Results and Discussion' followed by a section `Conclusion'. Please refer to Journal-level guidance for any specific requirements. 

% \bigskip

% Springer journals and proceedings: \url{https://www.springer.com/gp/editorial-policies}

% Nature Portfolio journals: \url{https://www.nature.com/nature-research/editorial-policies}

% \textit{Scientific Reports}: \url{https://www.nature.com/srep/journal-policies/editorial-policies}

% BMC journals: \url{https://www.biomedcentral.com/getpublished/editorial-policies}
% \end{flushleft}

\begin{appendices}

%%=============================================%%
%% For submissions to Portfolio Journals %%



\end{appendices}


% \nocite{*}

%Any of the below method can be used for citing the source of resources used.
% \bibliographystyle{nature}
\bibliography{ml4h-bibliography.bib} % Use the/ bibliography.bib file as the source of references

\begin{thebibliography}{10}

\bibitem{calvo2009upstream}
S.~E. Calvo, D.~J. Pagliarini, and V.~K. Mootha.
\newblock Upstream open reading frames cause widespread reduction of protein expression and are polymorphic among humans.
\newblock {\em Proceedings of the National Academy of Sciences}, 106(18):7507--7512, 2009.

\bibitem{constant2023deep}
D.~A. Constant, J.~M. Gutierrez, A.~V. Sastry, R.~Viazzo, N.~R. Smith, J.~Hossain, D.~A. Spencer, H.~Carter, A.~B. Ventura, M.~T. Louie, et~al.
\newblock Deep learning-based codon optimization with large-scale synonymous variant datasets enables generalized tunable protein expression.
\newblock {\em BioRxiv}, pages 2023--02, 2023.

\bibitem{earle2021evidence}
K.~A. Earle, D.~M. Ambrosino, A.~Fiore-Gartland, D.~Goldblatt, P.~B. Gilbert, G.~R. Siber, P.~Dull, and S.~A. Plotkin.
\newblock Evidence for antibody as a protective correlate for covid-19 vaccines.
\newblock {\em Vaccine}, 39(32):4423--4428, 2021.

\bibitem{elnaggar2021prottrans}
A.~Elnaggar, M.~Heinzinger, C.~Dallago, G.~Rehawi, Y.~Wang, L.~Jones, T.~Gibbs, T.~Feher, C.~Angerer, M.~Steinegger, et~al.
\newblock Prottrans: Toward understanding the language of life through self-supervised learning.
\newblock {\em IEEE transactions on pattern analysis and machine intelligence}, 44(10):7112--7127, 2021.

\bibitem{fallahpour2024codontransformer}
A.~Fallahpour, V.~Gureghian, G.~J. Filion, A.~B. Lindner, and A.~Pandi.
\newblock Codontransformer: a multispecies codon optimizer using context-aware neural networks.
\newblock {\em bioRxiv}, pages 2024--09, 2024.

\bibitem{fu2020codon}
H.~Fu, Y.~Liang, X.~Zhong, Z.~Pan, L.~Huang, H.~Zhang, Y.~Xu, W.~Zhou, and Z.~Liu.
\newblock Codon optimization with deep learning to enhance protein expression.
\newblock {\em Scientific reports}, 10(1):17617, 2020.

\bibitem{gong2023integrated}
H.~Gong, J.~Wen, R.~Luo, Y.~Feng, J.~Guo, H.~Fu, and X.~Zhou.
\newblock Integrated mrna sequence optimization using deep learning.
\newblock {\em Briefings in Bioinformatics}, 24(1):bbad001, 2023.

\bibitem{goulet2023codon}
D.~R. Goulet, Y.~Yan, P.~Agrawal, A.~B. Waight, A.~N.-s. Mak, and Y.~Zhu.
\newblock Codon optimization using a recurrent neural network.
\newblock {\em Journal of Computational Biology}, 30(1):70--81, 2023.

\bibitem{jain2023icor}
R.~Jain, A.~Jain, E.~Mauro, K.~LeShane, and D.~Densmore.
\newblock Icor: improving codon optimization with recurrent neural networks.
\newblock {\em BMC bioinformatics}, 24(1):132, 2023.

\bibitem{kim2022modifications}
S.~C. Kim, S.~S. Sekhon, W.-R. Shin, G.~Ahn, B.-K. Cho, J.-Y. Ahn, and Y.-H. Kim.
\newblock Modifications of mrna vaccine structural elements for improving mrna stability and translation efficiency.
\newblock {\em Molecular \& cellular toxicology}, pages 1--8, 2022.

\bibitem{kim2024computational}
Y.-A. Kim, K.~Mousavi, A.~Yazdi, M.~Zwierzyna, M.~Cardinali, D.~Fox, T.~Peel, J.~Coller, K.~Aggarwal, and G.~Maruggi.
\newblock Computational design of mrna vaccines.
\newblock {\em Vaccine}, 42(7):1831--1840, 2024.

\bibitem{kudla2006high}
G.~Kudla, L.~Lipinski, F.~Caffin, A.~Helwak, and M.~Zylicz.
\newblock High guanine and cytosine content increases mrna levels in mammalian cells.
\newblock {\em PLoS Biology}, 4(6):e180, 2006.

\bibitem{lecun2015deep}
Y.~LeCun, Y.~Bengio, and G.~Hinton.
\newblock Deep learning.
\newblock {\em nature}, 521(7553):436--444, 2015.

\bibitem{lorenz2011viennarna}
R.~Lorenz, S.~H. Bernhart, C.~H{\"o}ner~zu Siederdissen, H.~Tafer, C.~Flamm, P.~F. Stadler, and I.~L. Hofacker.
\newblock Viennarna package 2.0.
\newblock {\em Algorithms for Molecular Biology}, 6:1--14, 2011.

\bibitem{nedialkova2015optimization}
D.~D. Nedialkova and S.~A. Leidel.
\newblock Optimization of codon translation rates via trna modifications maintains proteome integrity.
\newblock {\em Cell}, 161(7):1606--1618, 2015.

\bibitem{parvathy2022codon}
S.~T. Parvathy, V.~Udayasuriyan, and V.~Bhadana.
\newblock Codon usage bias.
\newblock {\em Molecular biology reports}, 49(1):539--565, 2022.

\bibitem{ranaghan2021assessing}
M.~J. Ranaghan, J.~J. Li, D.~M. Laprise, and C.~W. Garvie.
\newblock Assessing optimal: inequalities in codon optimization algorithms.
\newblock {\em BMC biology}, 19:1--13, 2021.

\bibitem{ren2024codonbert}
Z.~Ren, L.~Jiang, Y.~Di, D.~Zhang, J.~Gong, J.~Gong, Q.~Jiang, Z.~Fu, P.~Sun, B.~Zhou, et~al.
\newblock Codonbert: a bert-based architecture tailored for codon optimization using the cross-attention mechanism.
\newblock {\em Bioinformatics}, page btae330, 2024.

\bibitem{sharp1987codon}
P.~M. Sharp and W.-H. Li.
\newblock The codon adaptation index-a measure of directional synonymous codon usage bias, and its potential applications.
\newblock {\em Nucleic acids research}, 15(3):1281--1295, 1987.

\bibitem{teo2022review}
S.~P. Teo.
\newblock Review of covid-19 mrna vaccines: Bnt162b2 and mrna-1273.
\newblock {\em Journal of pharmacy practice}, 35(6):947--951, 2022.

\bibitem{vitiello2021brief}
A.~Vitiello and F.~Ferrara.
\newblock Brief review of the mrna vaccines covid-19.
\newblock {\em Inflammopharmacology}, 29(3):645--649, 2021.

\bibitem{welch2009you}
M.~Welch, A.~Villalobos, C.~Gustafsson, and J.~Minshull.
\newblock You're one in a googol: optimizing genes for protein expression.
\newblock {\em Journal of the Royal Society Interface}, 6(suppl\_4):S467--S476, 2009.

\bibitem{yang2019generative}
D.~K. Yang, S.~L. Goldman, E.~Weinstein, and D.~Marks.
\newblock Generative models for codon prediction and optimization.
\newblock {\em Machine Learning in Computational Biology}, 2019.

\bibitem{zhang2023algorithm}
H.~Zhang, L.~Zhang, A.~Lin, C.~Xu, Z.~Li, K.~Liu, B.~Liu, X.~Ma, F.~Zhao, H.~Jiang, et~al.
\newblock Algorithm for optimized mrna design improves stability and immunogenicity.
\newblock {\em Nature}, 621(7978):396--403, 2023.

\end{thebibliography}

% %% if required, the content of .bbl file can be included here once bbl is generated
% %%\input ml4h-article.bbl

% \input ml4h-sample-bib.tex

\end{document}